# Enhanced dielectric breakdown performance of anatase and rutile titania based nano-oils

**Ajay Katiyar [a, b, *], Purbarun Dhar [b, #], Tandra Nandi [c, ×],**

**Lakshmi Sirisha Maganti [b] and Sarit K. Das [b, $]**

[a] Research and Innovation Centre (DRDO), Indian Institute of Technology Madras Research Park,
Chennai–600 113, India

[b] Department of Mechanical Engineering, Indian Institute of Technology Madras,
Chennai–600 036, India

[c] Defence Materials and Stores Research and Development Establishment (DRDO), G.T. Road,
Kanpur–208 013, India

*Corresponding author: Electronic mail: ajay_cim@rediffmail.com

*Tel No: +91-44-22548-222

*Fax: +91-44-22548-215

$Corresponding author: Electronic mail: skdas@iitm.ac.in

$Phone: +91-44-2257 4655

$Fax: +91-44-2257 4650

# E-mail: pdhar1990@gmail.com

× E-mail: tandra_n@rediffmail.com

# Abstract

Nano–oils synthesized by dispersing dielectric nanostructures counter common intuition as such nano–oils possess substantially higher positive dielectric breakdown voltage with reduced streamer velocities than the base oils. Nano–oils comprising stable and dilute homogeneous dispersions of two forms of titanium (IV) oxide ($TiO_2$) nanoparticles (Anatase and Rutile) have been experimentally examined and observed to exhibit highly enhanced dielectric breakdown strength compared to conventional transformer oils. In–depth survey of literature yields that research on enhancing insulation properties of mineral oils by utilizing anatase and rutile titania nanoparticles is nought. The present study involves titania dispersed in two different grades of transformer oils, both with varied levels of thermal treatment, to obtain consistent and high degrees of enhancement in the breakdown strength, as well as high degrees of increment in the survival of the oils at elevated electrical stressing compared to the base oils, as obtained via detailed twin parameter Weibull distribution analysis of the experimental observations. The experimental results demonstrate augmentation in the breakdown strength ~79 % enhancement for Anatase and ~51% Rutile at relatively low concentrations of 0.1 wt. %. It is also observed that heat treatment of the nano–oils further enhances the dielectric breakdown performance. Additionally, further enhancements are observed at elevated operational temperatures. Detailed studies on the performance of the nano–oils in presence of variant quantities of moisture have also been put forward. The grossly different charging dynamics of dielectric nanoparticles localized within the oil has been proposed to be responsible for efficient electron trapping, leading to decrease in the positive streamer velocity and resulting in high dielectric breakdown voltage. The differences in the performance of anatase and rutile has been explained based on the electronic structure of the two and the affinity towards electron scavenging and the theory has been found to validate the experimental observations. The present study can find prime implications and promise in the field of dielectrics and liquid insulation of power electronics and electrical devices.

# 1. Introduction

Insulating materials play a major role in the design, operation and performance of electrical devices, especially those involving high voltages or high power densities. Mineral oils are the most preferred coolant and insulators in transformers due their cooling potential, high thermal stability and excellent electrical insulation properties. Insulation is provided to isolate the inside windings, prevent leakage of field flux lines as well as cool the transformer core and maintain a stable operating temperature. A major focus of present day research in the field of insulating fluid rests on increasing the insulating strength of the liquids without compromising its cooling and anti–corrosion properties, an avenue where seeding with dielectric nanostructures has emerged as a potential solution. In recent times, various smart applications and abilities of nanostructures in uniform and stable colloidal phases have been investigated [1–5]. Dielectric breakdown (BD) of insulating oil is one of the key characteristics that govern safe operation of modern electrical power systems which are cooled by such oils, such as transformers, large scale capacitors and resonator coils, etc. Although an area of practice, usage of chemical additives to enhance dielectric breakdown performance of oil have their own limitations, such as corrosion of components, possible reaction at high electric field intensities and temperatures, possibility of flammability in case of corona discharge at high voltages and so forth. Consequently, in recent times, researchers have started looking for alternatives to such chemical additives in the form of nanostructures [6, 7] and have also demonstrated that graphene and CNT based transformer nano–oils demonstrate highly augmented BD strength at very dilute concentrations [8]. The dielectric strength of insulating oils; especially transformer oils, has been studied comprehensively and various dielectric breakdown mechanisms have been proposed widely [9–15] and the relationships between life cycle of transformers and thermal performance of the insulating oils under variant load conditions has also been reported [16–18]. In general, the chemical and dielectric properties of transformer oils are prone to irreversible changes due to aging. Variant factors such as high temperature, dissolved gases, humidity and frequent electrical discharge may accelerate the aging process [19, 20], leading to uncertainty in comprehending the system, which is why statistical analyses have been extensively utilised to quantitatively describe the BD failure [21–23].

Moisture is soluble more or less to certain extent in all types of insulating oils; hence the existence of moisture in the oil leads to reduction in dielectric breakdown strength and increase in dielectric losses [24]. The presence of moisture further reduces the effective dielectric strength of the oils when it separates from oil and deposits on the electrodes or the live terminals [25]. The dielectric strength has been reported to enhance as a function of temperature and also depends upon the grade and viscosity of the oils [26]. Various nanostructures have been utilized to formulate nanofluids; especially alumina in the form of nanopowder exhibits the ability to absorb water molecules from the surrounding environment and given its stature as an active and corrosive material [27], it can enhanced water content in the oil as well as corrode metallic components. Titanium dioxide is a widely available high dielectric nanomaterial and can be utilized for a variety of applications in the form of nanofluids, colloids or in combination with mineral oils to enhance their capability to store thermal energy as well as modify wear or frictional resistance [28–30]. $TiO_2$ is obtained naturally in three major crystalline structures viz. rutile, anatase and brookite and the physical and chemical properties are dependent on the type of phase [31]. Presence of $TiO_2$ nanostructures is expected to influence the distribution of electric field lines within matter due to the high relative permittivity and the crystal structure of the former. The dispersion of nanostructures in transformer oils has been reported to enhance the dielectric performance by capturing the fast streamer electrons and transforming the negatively charged but slower electrons accordingly [7, 32]. Such nanostructures can thus, in a suspension, be expected to tune the material dielectric properties of the liquid phase, leading to modified BD characteristics.

## 2. Experimental Section

### 2.1. Instrumentation

#### 2.1. a. Liquid Dielectric Breakdown Device

Investigating the dielectric breakdown (BD) strength of the conventional transformer oils (T.O) and the nano–oils (N.O) requires an accurate device to analyse or measure the breakdown voltage in accordance to the specified ASTM/IEC standards and codes. For the present study, an automated high voltage (HV) device has been used and components of the same have been demonstrated in Fig. 1 (a). The device consists of two main sections, viz. test basin (as demonstrated in Fig. 1 (a3)) with HV terminals (as demonstrated in Fig. 1 (a1)) and the auto–transformer. The device is operated with a single phase 220 V AC input and utilizes the in–built HV transformer to step up the input. The HV is supplied across the electrode terminals of the test basin filled with the test fluid and the current flowing across the terminals is read by a precise ammeter. At the moment of BD, an arc of high current density is discharged across the electrodes (as demonstrated in Fig. 1 (c) and (d)) and the controller trips the main circuit. The voltmeter reading at the tripping point is read as the breakdown voltage of the test fluid. The experiments have been executed in accordance with the ASTM D–877 utilizing two hemispherical electrodes of 5 mm diameter (as demonstrated in Fig. 1 (a2)). The distance between the two electrodes has been accordingly maintained fixed at 5 mm and the rise of the HV across the terminals is controlled automatically at an approximate rate of 2 kV/s. Experiments have also been performed to observe the effects of operating temperature, heat treatment and moisture on the DB strength of the N.O.s. The heat treatment protocol involves the oil being heated to elevated temperatures (~100 °C) in an oven and cooled undisturbed to room temperature under vacuum environment and the experiments are carried out thereafter.

## 2.1. b. Digital Moisture Titration Unit

The presence of moisture in the oil is experimentally measured by utilizing a digitized Karl Fischer titration unit (MKS 520, Japan), as demonstrated in Fig. 1 (b). Its main components comprise a reference sample cup, digital moisture meter and the titration facility. In order to measure moisture level in the oil, an initial calibration pre–titration is carried out on the reference/standard sample provided by the manufacturer. Subsequently, the main sample (30 ml) is loaded in a syringe and the weight of the syringe with the sample is measured with a precision balance before injecting the sample in to the sample cup. The weight of the empty syringe is measured and titration is performed run for the sample. Measured weights are used to compute the moisture content in oil in part per mole (ppm). The measured values of moisture content in the two grades of oil used, namely, oil 1 and oil 2 is ~26 and 12 ppm respectively at room temperature.

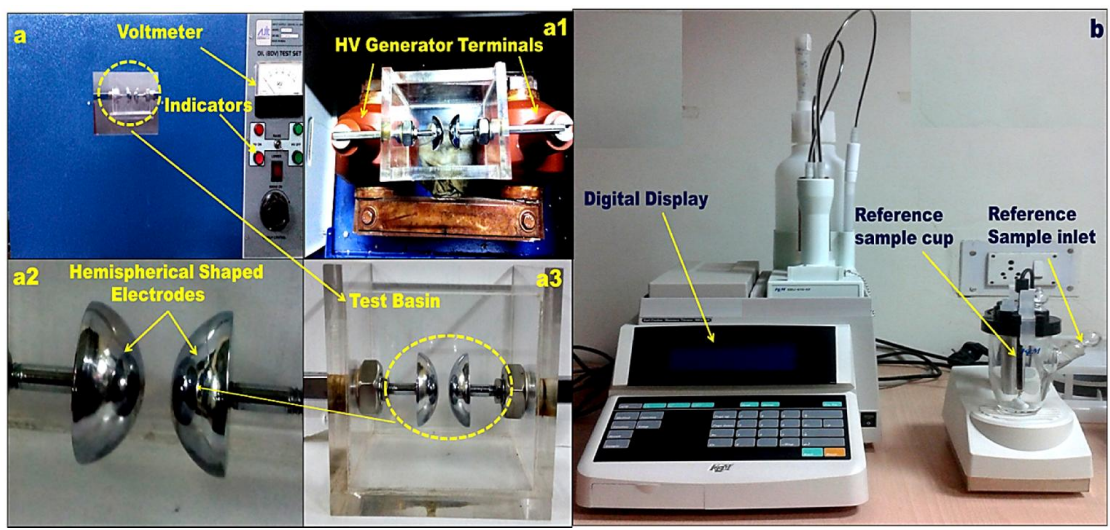

**Figure 1:** Components of the liquid dielectric breakdown device **(a)** Front view of the test setup. The indicators exhibit the status, either on or off, of the HV terminals across the sample. The system is automated to increase or decrease the voltage across the terminals (displayed by the voltmeter) at the desired rate. **(a1)** Top view of the test setup showing the HV generator coils and the test basin with protruding ends of the electrodes rested on the terminals. **(a2)** Front view of test basin with the electrode assembly. **(a3)** The hemi–spherical electrodes maintained at fixed gap. **(b)** The Karl Fischer titration unit to quantify the moisture content in the oil/liquid samples.

## 2.2. Materials and characterization

The nanoparticles (NPs) utilized in the present study have been procured from Nanoshel Inc. India (> 99.5 % purity as per suppliers data) and dried at temperature over a range of ~ 90–100 °C in an oven for 2–3 hours before use to rid the particles of moisture. Oleic acid (99.98 % purity), utilized as a stabilizer, has been procured from Sisco Research Labs (Mumbai, India). The physical appearance and size of the materials have been determined by High Resolution Transmission Electron Microscopy (HRTEM). Initially, the anatase and rutile $TiO_2$ NPs were dispersed in acetone individually by utilizing a probe type ultrasonicator so as to break the agglomerations and form a homogeneous and stable suspension. A single drop of the sample was taken on the carbon coated copper grid for the TEM. Fig. 2 (a) and (b) demonstrate the TEM images of the rutile and anatase $TiO_2$ NPs respectively. The average dimensions of the $TiO_2$ NPs is found to be over a range of 15–20 nm and 20–30 nm, for the rutile (oblate and spherical particles) and anatase (predominantly cubic and polyhedral particles) NPs, respectively.

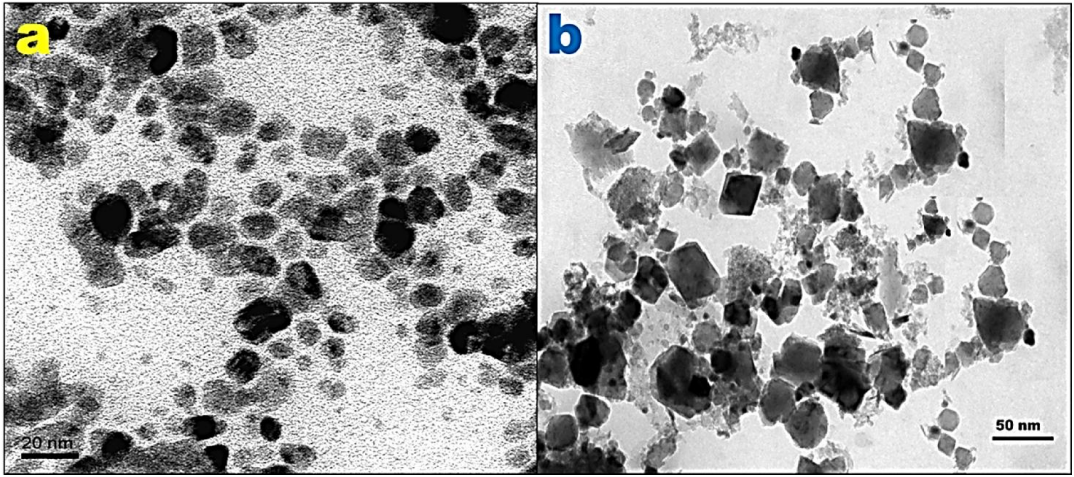

**Figure 2:** Transmission electron microscopy (TEM) of NPs: **(a)** TiO$_2$ (rutile) NPs with mean diameter of particles over a range of 15–25 nm. **(b)** TiO$_2$ (anatase) NPs with average dimension of particles over a range of 20–30 nm.

## 2.3 Insulating Nano-oils

Two grades of T.Os have been utilized and have been represented as Oil 1 (without thermal treatment) and Oil 2 (heated up to 100 °C in an oven and then cooled to room temperature under vacuum condition) hereafter. N.Os have been synthesized by infusing the prerequisite amount of NPs over a range of 0.01–0.3 wt. %, followed by ultra–sonication for 2–3 hrs by utilizing a probe type ultrasonicator. Oleic acid (OA) has been used as the stabilizing or capping agent to improve the stability and shelf life of the nanosuspensions. The amount of OA utilized is varied from 0.2–3 mL per 500 mL of oil, depending upon the concentration of NPs. As observed from experiments, the existence of OA in the oils in such dilute proportions has insignificant effect on the viscosity, thermal conductivity and breakdown strength of the T.O. Images of the T.O and TiO$_2$ diffused N.Os samples have been illustrated in Figs. 3 (a) and (b) respectively and the physical properties of the untreated and the heat treated T.Os have been tabulated in Table 1. The flash and fire points for the oils are measured, employing a Pensky Martens oil testing apparatus.

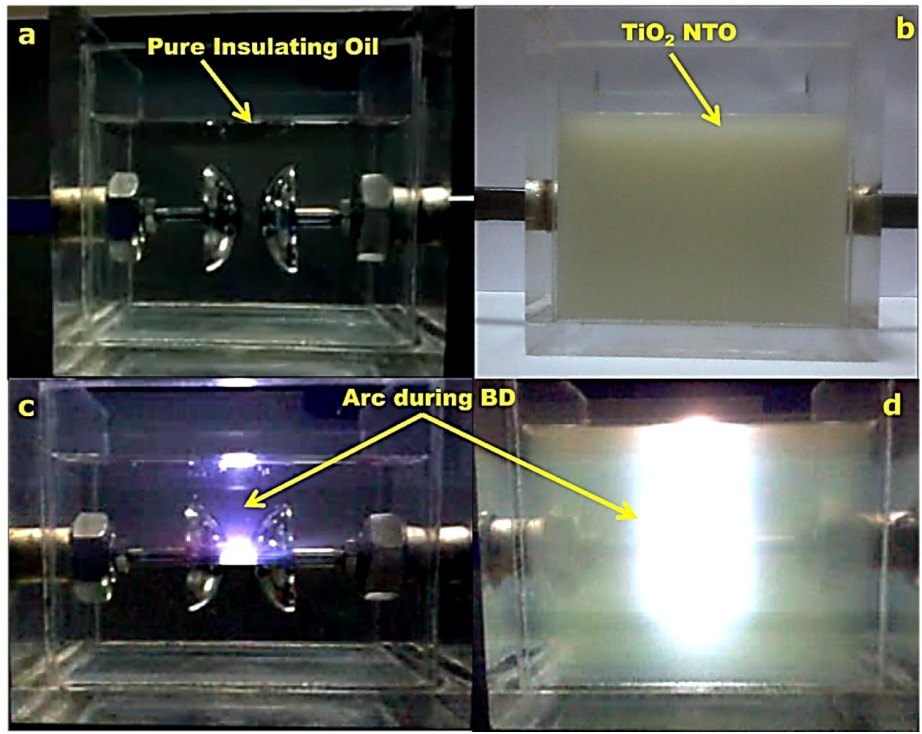

**Figure 3:** **(a)** The base T.O **(b)** TiO$_2$ diffused NO. (c) The arc generated during the BD failure of T.O and **(d)** TiO$_2$ infused NO (the images of arc have been captured for 0.05 wt. % of NPs).

**Table 1:** Physical properties of the T.Os

| Parameters at 25°C | ~ Value |
|---|---|
| Flash point (°C) | Oil1–146 |
| | Oil2–151 |
| Fire point (°C) | Oil1–156 |
| | Oil2–162 |
| Density (kg/m$^3$) | Oil1–870 |
| | Oil2–807 |
| BD voltage (kV/mm) | Oil1–7.0 |
| | Oil2–8.0 |
| Viscosity (mPas) | Oil1–16.0 |
| | Oil2–18.2 |

# 3.    Results and Discussion

### 3.1 Enhanced dielectric breakdown strength of NOs

Experimental results conclusively demonstrate that the presence of dielectric NPs in T.O. leads to enhancement of the dielectric performance of the oil. Oil 1 and Oil 2 have been observed to possess BD strength of ~32 and 40 kV respectively, when tested with the standard 5 mm electrode gap configuration and this value shows no appreciable deviation when trace quantities of OA is added. The BD failure of the insulating oil occurs with a discharge phenomenon across the gap, generated due to the high current passing through the streamer tunnel within the nonconductive medium. The BD characteristics depend upon the size of arc discharge as well as thermionic emission and field emission of electron from the electrodes supporting the arc. Fig. 3 (c) and (d) demonstrate the HV arc generated during the failure of T.O and N.O and it has been observed that in general, the arc characteristics such as number of intermittent arcs and noise generated due to the discharge are lower in intensity in case of the N.O.s; thereby providing qualitative evidence on the enhanced device safety in case of N.O.s.

The BD characteristics of Rutile $TiO_2$ based N.Os (Rutile oil 1 and Rutile oil 2) as a function of NPs concentration has been illustrated in Fig. 4 (a) and the corresponding magnitudes of enhancement and the critical concentration at which maximum augmentation is attained have been illustrated in Fig. 4 (b). It is observed that the presence of $TiO_2$ in the T.O leads to higher insulting properties with the BD voltage shooting up to a maximum of ~57 and 62.5 kV (at 0.05 wt. % concentrations) for Rutile oil 1 and Rutile oil 2 respectively and the magnitudes of the corresponding enhancement attained for the same are ~ 64 and 56 %, respectively. Similarly, the maximum BD voltages for Anatase oil1 and Anatase oil 2 are found to be ~ 66 and 73 kV (at 0.15 wt. % concentrations) respectively and have been illustrated in Fig. 5(a). The magnitudes of the corresponding enhancement for the same are observed as ~ 88 and 82 % respectively (as illustrated in Fig. 5 (b)). Anatase based N.Os demonstrates superior BD strengths compared to the rutile N.O.s but at relatively higher concentration of NPs. It is observed that all the tested samples exhibit enhancement in BD strength up to a critical concentration of NPs and a subsequent drop in BD strength occurs [8].

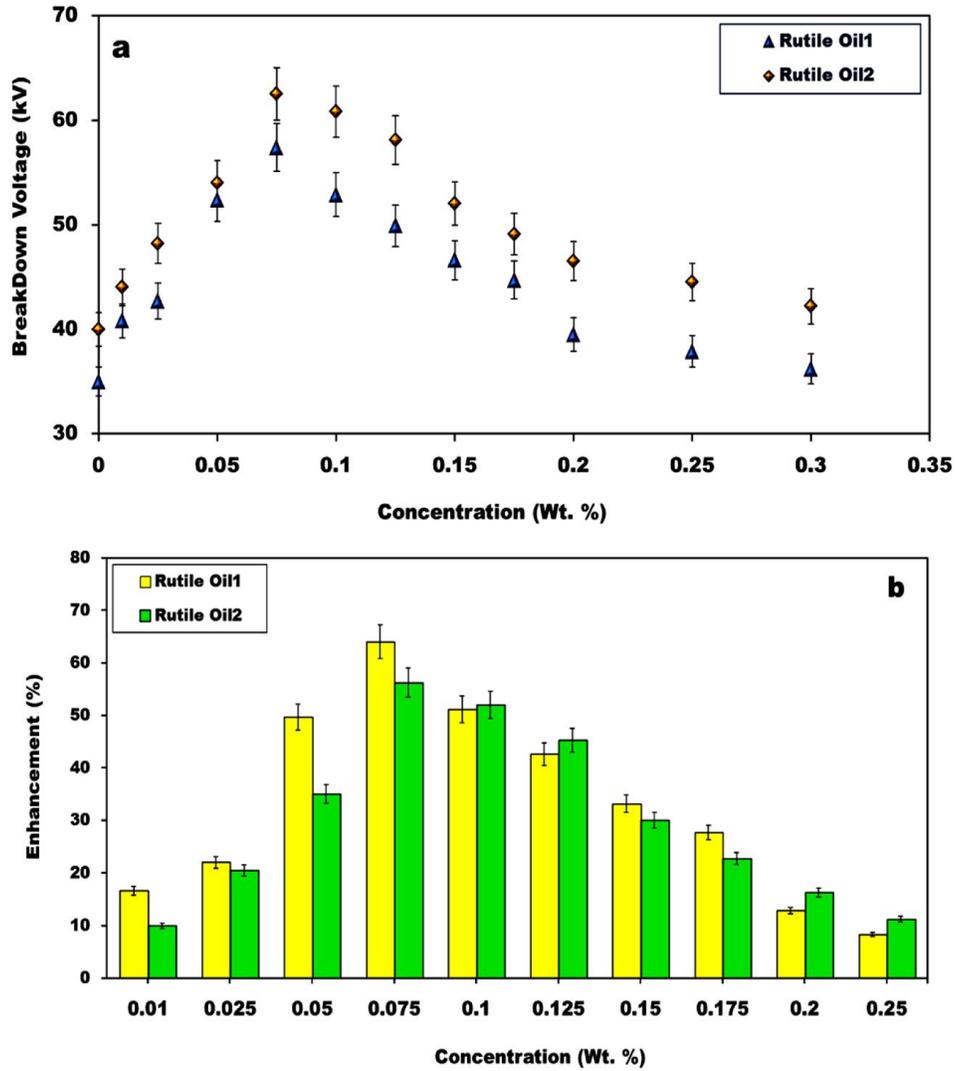

**Figure 4: (a)** Dielectric breakdown characteristics of Rutile TiO$_2$ infused NOs (Rutile oil 1 and Rutile oil 2) as a function of NPs concentration. **(b)** The corresponding magnitudes of percentage enhancement and the critical concentration at which maximum BD strength occurs at room temperature.

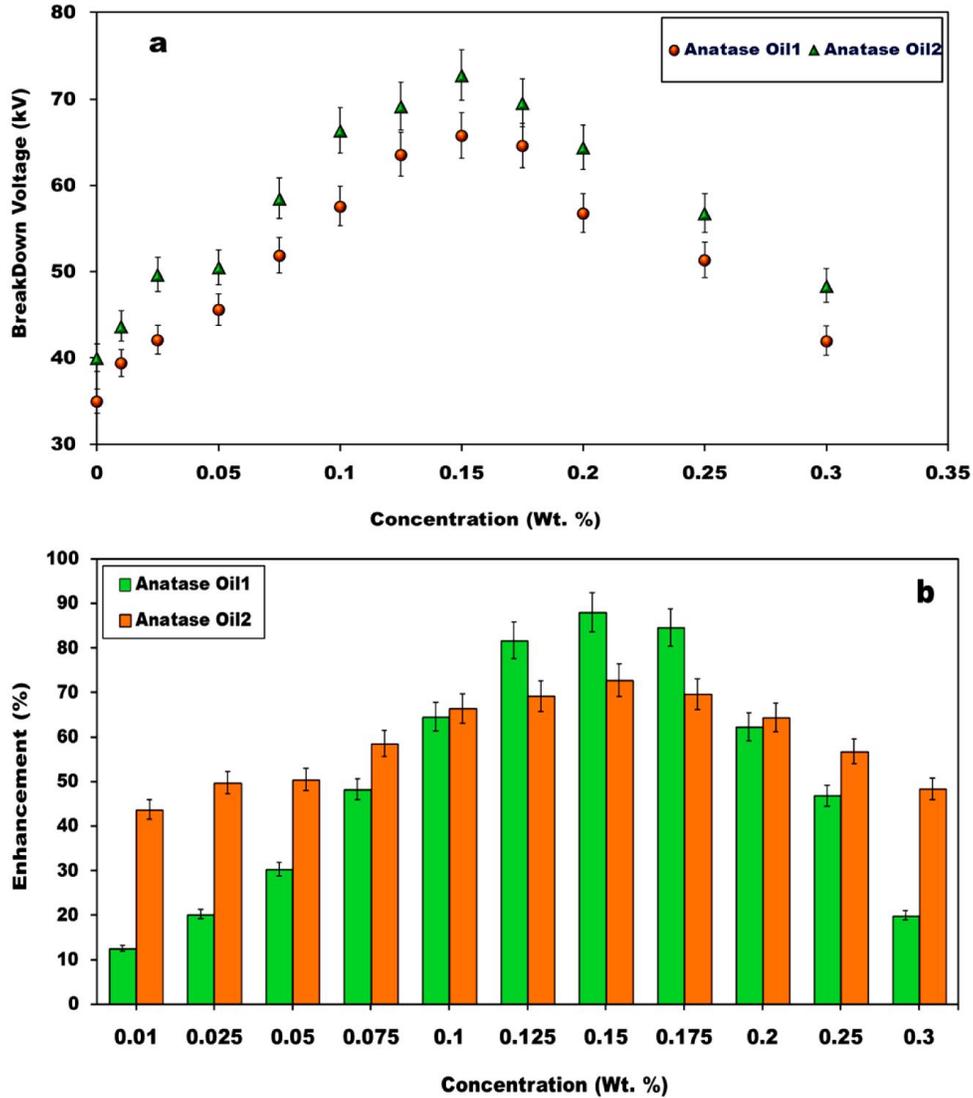

**Figure 5: (a)** Dielectric breakdown characteristics of anatase TiO$_2$ infused NOs (Anatase oil 1 and Anatase oil 2) as a function of NPs concentration **(b)** The corresponding magnitudes of percentage enhancement and critical particle concentration at which maximum DB voltage occurs at room temperature.

The enhancement of BD voltage of T.O due to addition of dielectric NPs is caused by the modification of the fundamental electro–hydrodynamic transport routes within the N.O. when exposed to high voltage traumas. The higher BD voltage is a result of the changed electrodynamics of the streamers within the oil due to the presence of nanostructures. The temporal growth of the zones of ionized molecules (streamers) that move towards the opposite electrode is governed by the BD strength of the liquid phase. The growth dynamics of the streamers is directly proportional to the BD strength as well as the applied voltage and the nature of the field in space. The presence of NPs with high dielectric constant is responsible for changing the localized permittivity of the oil, leading to decrease in the positive streamer propagation velocity and delaying breakdown event. Since the relative permittivity of the nanomaterial is much higher than that of the oil, the field lines converge onto the nanomaterials, causing the streamers to be directed on to the nanostructures, which then efficiently scavenge the same before they can travel to the opposite electrode. Thus if the response period of the ionized constituents responsible for development of the streamers can be enhanced, the possibility of BD for the liquid at a stated voltage is axiomatically reduced. However, as far as the following study is concerned, the differences in DB strengths observed for the anatase and rutile based oils requires a reasoning based on the physical properties of the two phases of $TiO_2$ as the particles exhibit similar size distributions.

The fact that scavenging of electrons is the established theory by which nanostructures delay DB phenomenon [8], a proper understanding of the scavenging efficacy of the two types of nanomaterial should be looked into. This may be explained qualitatively based on the electronic structure of the constituent surface atoms (the members active in charge scavenging) of the nanoparticles. The energy gap between the conduction and the valence bands for the different types of particles can be argued to play an important role in the charge scavenging capabilities of the nanoparticles. Studies reveal that the band gaps of anatase and rutile $TiO_2$ are 3.2 and 3.03 eV respectively, indicating that the gap in anatase is higher by ~ 0.2 eV compared to the rutile [33]. During scavenging, the stray electrons are adsorbed on to the surface atoms, i.e. they lodge themselves in the outermost shells of the electron structure of the titania atoms. The higher band gap in anatase indicates that its atoms have larger capability to absorb the incoming electrons due to lesser repulsion from the valence band electron population. This repulsion is, in general, greater in case of rutile as the lower band gap implies that the valence band electrons have larger propensity to travel to the conduction band by utilizing the energy absorbed due to impinging electron collisions. The qualitative illustration of the mechanism has been shown in Fig. 6. However, beyond a critical particle concentration, the inter–particle separation reduces to the point where the particles (indicated by green arrows in Fig. 7(b)), charged due to electron scavenging, lead to formation of conducting pathways for the streamers to tunnel through. The close vicinity of the charged nanostructures thereby act as agents that induce 'merger' of the two opposite steamers ahead of the predicted field strength and eventually lead to reduced breakdown strength (as illustrated in Fig. 7 (b)).

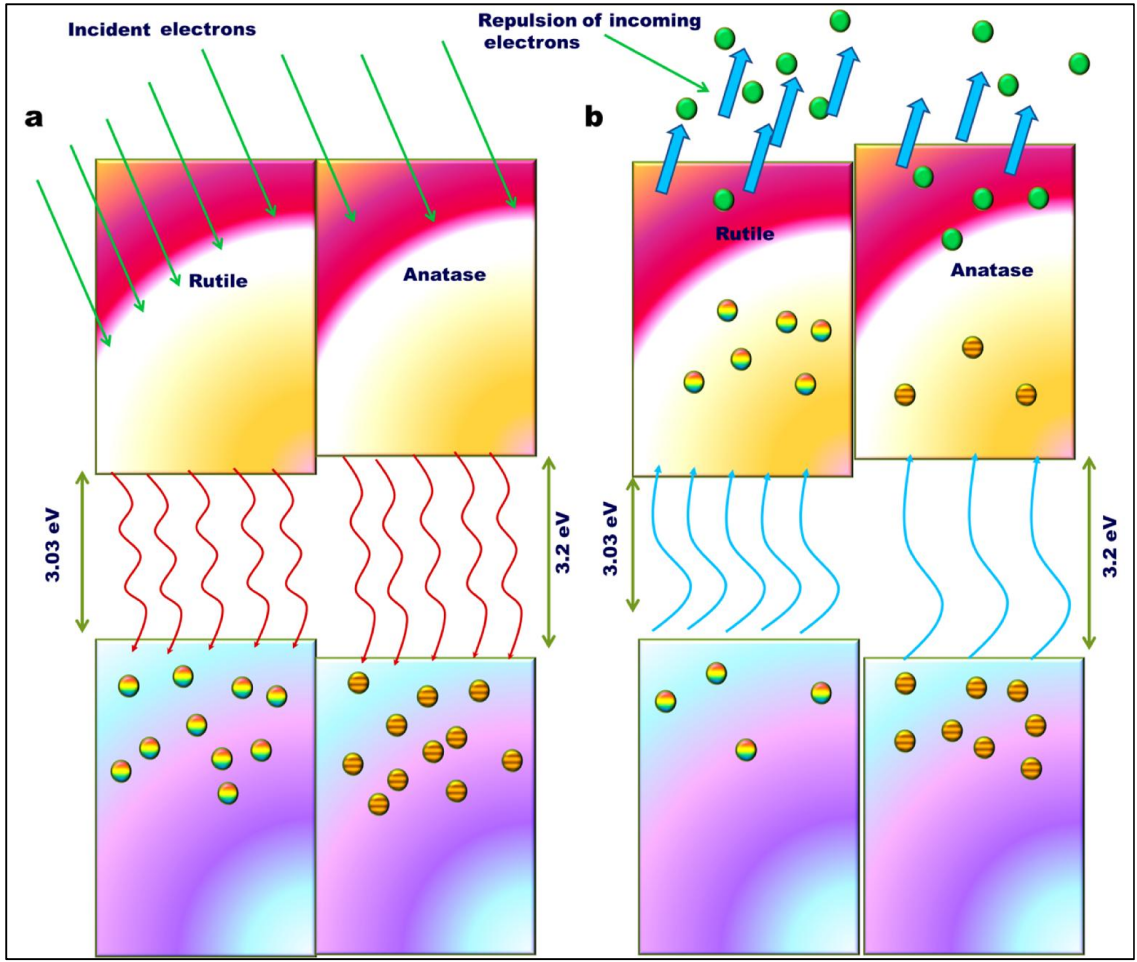

**Figure 6:** Illustration of the qualitative differences in breakdown strength induced to T.O. by anatase and rutile nanostructures. **(a)** The incident electrons (denoted by arrows) possessing high energies due to ionization are scavenged and lodged within the outer levels of the conduction band. These electrons eventually lose some parts of their energies to proceed to the lower energy states. The liberated energy is partly transferred (denoted by wavy arrows) to the valence band electrons. **(b)** A fraction of the energized valence electrons transit to the conduction band (denoted by curved arrows), increasing the effective electron density of the conduction band and leading to additional repulsion (denoted by thick arrows) to incoming electrons. Since the energy gap in anatase is higher, lesser number of electrons is expected to traverse the barrier, leading to lower electron density in the conduction band than rutile. This leads to reduced repulsion and hence anatase can scavenge more than its rutile counterpart.

The mechanism of the enhanced BD performance of dielectric NPs such as TiO$_2$ infused N.O.s can be described based on the induced electron trap depth corresponding to the maximum potential difference at the particle surface [34, 35]. The trap energy depth of the permanent dipole moment generated as a consequence in general insulating materials is observed to be $\sim$ 0.04–0.45 eV [36–37]. To induced trap depth can be evaluated from the expression for the generated potential onto the particle–fluid interface as (in spherical coordinates system) [38]:

$$V(r,\theta,\phi) = \frac{m\cos\phi}{\varepsilon_0\varepsilon_r 4\pi r^2}\sin\theta \tag{1}$$

Where, 'm' denotes the magnitude of the generated dipole moment and the expression for the same is expressible 'm=q$\delta$'. The 'q' indicates the induced electrical charge and '$\delta$' the dipole separation. When dielectric NPs are subjected to large electric fields (V), substantial electrical charge is induced on to the particle interface and the lines of flux converge on to the charged NPs. As a consequence, the positive and negative surface charge yields an electric dipole moment [38]. Since the average diameters of the TiO$_2$ NPs utilized in the present study exist over a range of 15–30 nm, $\delta$ may be expected equal to the average diameter (15–30 nm). Therefore the magnitude of $\delta$ is $\sim$ 10 fold greater than the distance between general chemical defects in the system ($\sim$ 1–2 Å). Hence, the dielectric NPs such as TiO$_2$ form excellent electron traps (due to convergence of the local field lines owing to the high dielectric constant) and enhances the BD strength of the N.Os. Smaller particles possess larger values of specific surface area allowable for scavenging and thus exhibit further enhanced DB strength in insulating fluids [39].

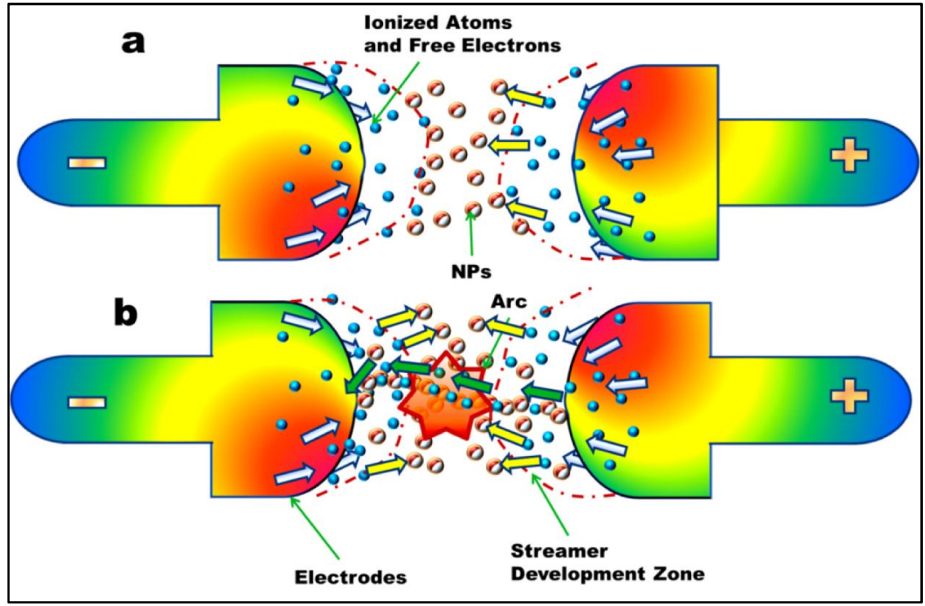

**Figure 7:** Mechanism of scavenging charge released due to ionization by the nanostructures near the electrodes (Light gray arrows indicate direction of traverse); **(a)** Dilute suspensions effectively capture the charges (indicated by yellow arrows) and delay the amalgamation of the two streamers, thereby enhancing breakdown strength. **(b)** Above a critical concentration, the inter–particle distance is reduced to the point wherein the liberated charges can traverse through the particle (indicated by green arrows), and 'flight' on the next particle due to the intimacy and/or direct contact and reach to the opposite streamer zone. The percolation chains thereby act as agents that induce 'merger' of the streamers, often leading to numerous intermittent arc discharge and eventually leading to reduced breakdown strength.

As the nanostructure population scavenges liberated electrons from the streamers, the average charge density per nanostructure approaches the saturation density and the inherent population is unable to scavenge further, and breakdown occurs. Consequently, with increasing population of nanostructures, i.e. with increased concentration, the amount of charge scavenged by the population before the average charge per nanostructure exceeds the saturation magnitude is higher than a low concentration case. As a result, increased concentration leads to enhanced magnitude of augmentation in breakdown voltage. However, although the enhancement is intuitively expected to be a linear function of growing concentration, in reality the enhancement follows a decaying growth rate, as evident from Fig. (5). Moreover, unlike $TiO_2$ and $Fe_3O_4$ based NOs [40, 41], the growth of enhancement saturates within dilute regimes.

The charging dynamics of the nanostructure population can be considered equivalent to the charging of a capacitive system, i.e. follows an exponentially decaying growth pattern. Ionization of fluid molecules leads to formation of both mobile free electrons as well as positively charged molecules. However, since the massive nature of the molecules lead to very low mobility compared to the electrons, the nanostructure come in contact with the free electrons more frequently than positive ions, leading to predominant scavenging of electrons. Consequently, the growth of negative charge on the nanostructures grow with time, which results in decaying charge scavenging ability in addition to the reduced potential of the nanostructure approaching its saturation point. With increasing population, the effective repulsion increases and thus the growth of enhancement decays out like a charging capacitor, to a concentration where the charge acceptance capability of the system reaches saturation. Since the electron density in anatase titanium oxide or iron oxide is very low, the resultant repulsion of inbound electrons is much smaller in magnitude at the same level of charging. As such, these systems can acquire charges up to greater concentrations and the growth of enhancement in such systems is often linear or bell shaped. However, NOs are incapable to enhance the breakdown strength beyond the critical concentration and decrease in enhancement is observed after the critical concentration.

Although mean values of DB strength indicates the increment in response, it does not provide data on the survival capabilities of the oil as far as any applied field intensity is concerned. The assurance of the degree of survival of the oil can be confirmed from statistics in terms of reliability. The twin parameter Weibull distribution is a good tool to model the BD reliability of dielectric materials for a wide range of operating voltages. The expression for the reliability for a material as obtained from the Weibull distribution is expressible as (Eqn. 2),

$$f(x;\alpha,\beta) = 1 - e^{-\left(\frac{x}{\beta}\right)^{\alpha}}$$ (2)

Where, 'x', 'α' and 'β' indicate the parameter under consideration (survival probability in the present case), the shape and the scale parameters respectively and the latter two are evaluated from the experimental datasets employing Weibull distribution analysis. Figure 8 (a) and (b) illustrate the reliability curves for oil 1, anatase oil 1 and rutile oil 1 samples at different concentrations and as function of operating voltages. It is observed that the reliability of the N.O.s is much more pronounced than the base oil for all concentrations up to the critical value. However, the range of reliability in case of rutile is relatively low for the optimum concentration (Fig. b) as compared to that of anatase, which indicates that although the mean strength of rutile based oils are higher, the survival probability of the anatase oils are higher at high fields. Similarly, the reliability plots for the samples oil 2, anatase oil 2 and rutile oil 2 have been demonstrated in Fig. 9. In general, it can be commented that the rutile based oils, while exhibiting higher degree of survival up to certain field strengths, experience more probability of breakdown beyond the mean value. In case of anatase, the probability of survival beyond the mean strength, although diminished, remains higher than the rutile oils of same concentration and for same applied field strength.

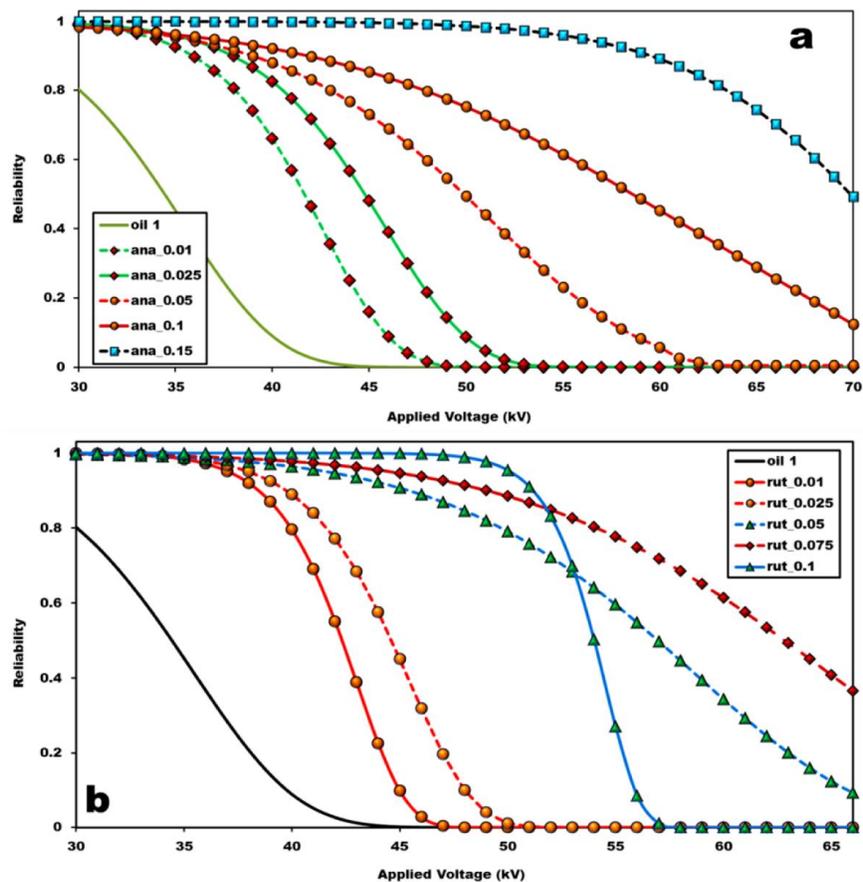

**Figure 8:** Reliability plots for the samples (a) Anatase oil 1 (b) Rutile oil 1 at different concentrations.

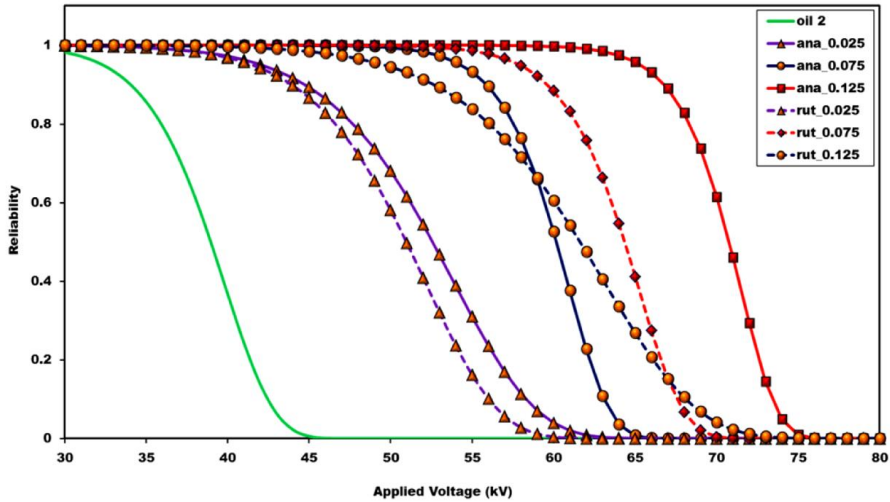

**Figure 9:** Probability plots for the samples oil 2, Anatase oil 2 and Rutile oil 2 at different concentrations.

### 3.2 Influence of temperature on the dielectric breakdown strength of NOs

To further understand the effect of operational parameters which influence the BD strength in real life scenarios, experiments have been carried out for temperature dependence of the BD strength. The influence of temperature on the BD characteristics of Rutile oil 1, Anatase oil 1 and oil 1 has been illustrated in Fig. 10 (a) and the magnitudes of corresponding enhancements in BD voltage as a function of temperature (compared to strength at room temperature) for the samples at critical NPs concentration have been illustrated in Fig. 10 (b). The temperature is varied over a range of 30–100 °C (since transformer cores are in general operational at ~ 80–100 °C) and is measured by a precision thermometer. Experimental results reveal that increase in temperature leads to further enhancement in the insulting properties of the base T.O compared to the insulating properties at room temperature. The samples have been chosen at the critical concentration for conducting the temperature study so as to scale the maximum possible enhancement under operating conditions.

The maximum magnitude of BD voltage is observed as ~ 47, 72.8 and 79.4 kV for oil 1, Rutile oil 1 and Anatase oil 1 respectively at 80 °C. The maximum magnitudes of the corresponding enhancements are ~ 35.7, 26.8, and 20.6 % are recorded as compared to the DB values at room temperature. The maximum BD voltage for oil 2, Rutile oil 2 and Anatase oil 2 is observed as ~ 51.5, 79.8 and 86.3 kV and at 90, 80 and 90 °C respectively (Fig. 11 (a)) and the maximum corresponding enhancement for the same is found to be ~ 28.7, 27.7 and 15 %, respectively (Fig. 11 (b)). The enhancement as a consequence of elevated temperature can be explained based on the enhanced thermal fluctuations of the nanostructures, leading to enhanced diffusion based scavenging capabilities of the nanoparticles. As the temperature increases, the nanostructures experience higher degrees of Brownian motion within the oil, leading to enhanced degree of chaotic motion within the oil. This effect, jointly with the increased instability of the charged entities due to lowering of fluidic viscous drag, leads to further deterioration of the stability of the growing streamers. The chaotic motion within the oil hampers the stable growth of the streamers and the enhanced thermo–diffusion leads to more effective scavenging, leading to delayed merger of the streamers [8]. The associated uncertainties in the experiments have been shown by corresponding error bars.

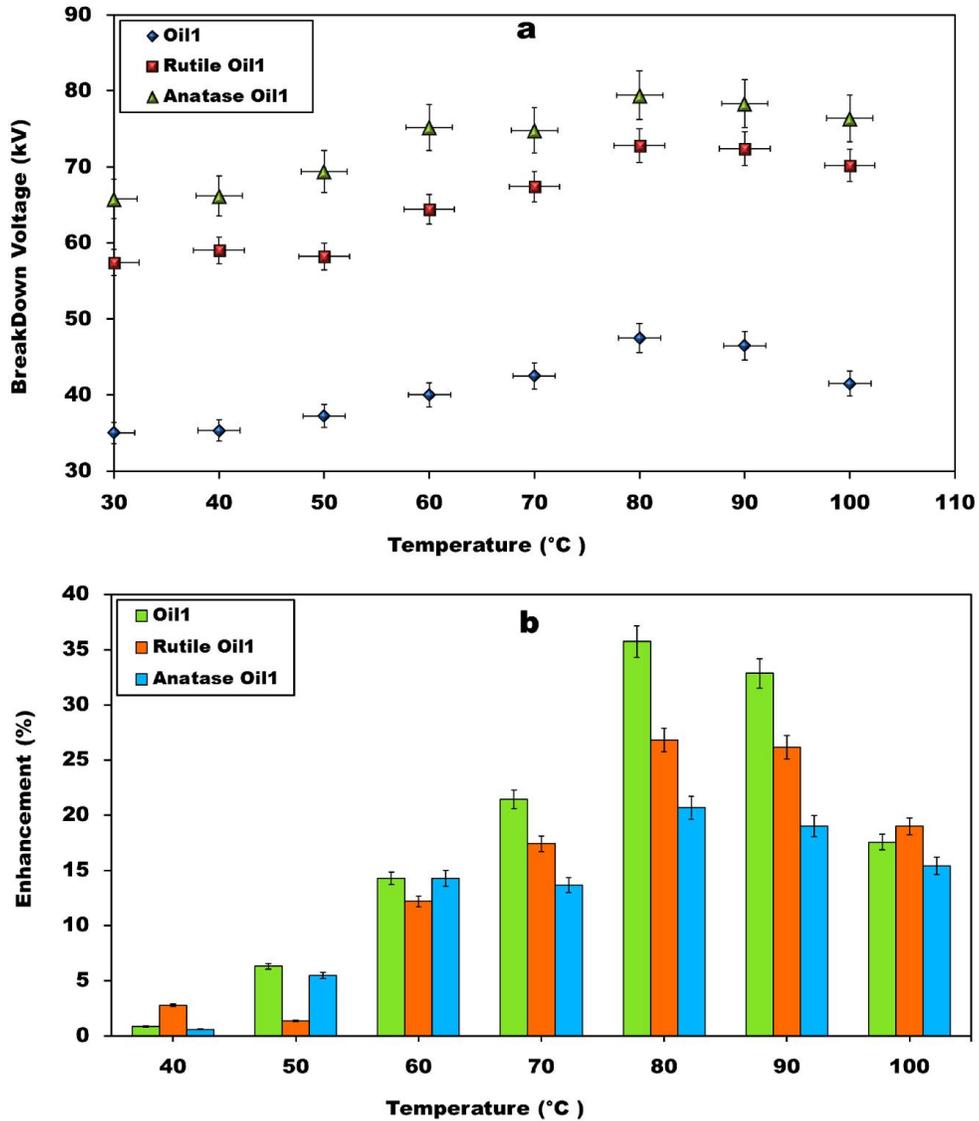

**Figure 10: (a)** The variation of dielectric breakdown voltage of N.O.s (Anatase oil 1 and Rutile oil 1) with respect to oil temperature **(b)** The corresponding magnitudes of percentage enhancement for BD voltage for the same, as compared to the performance at room temperature.

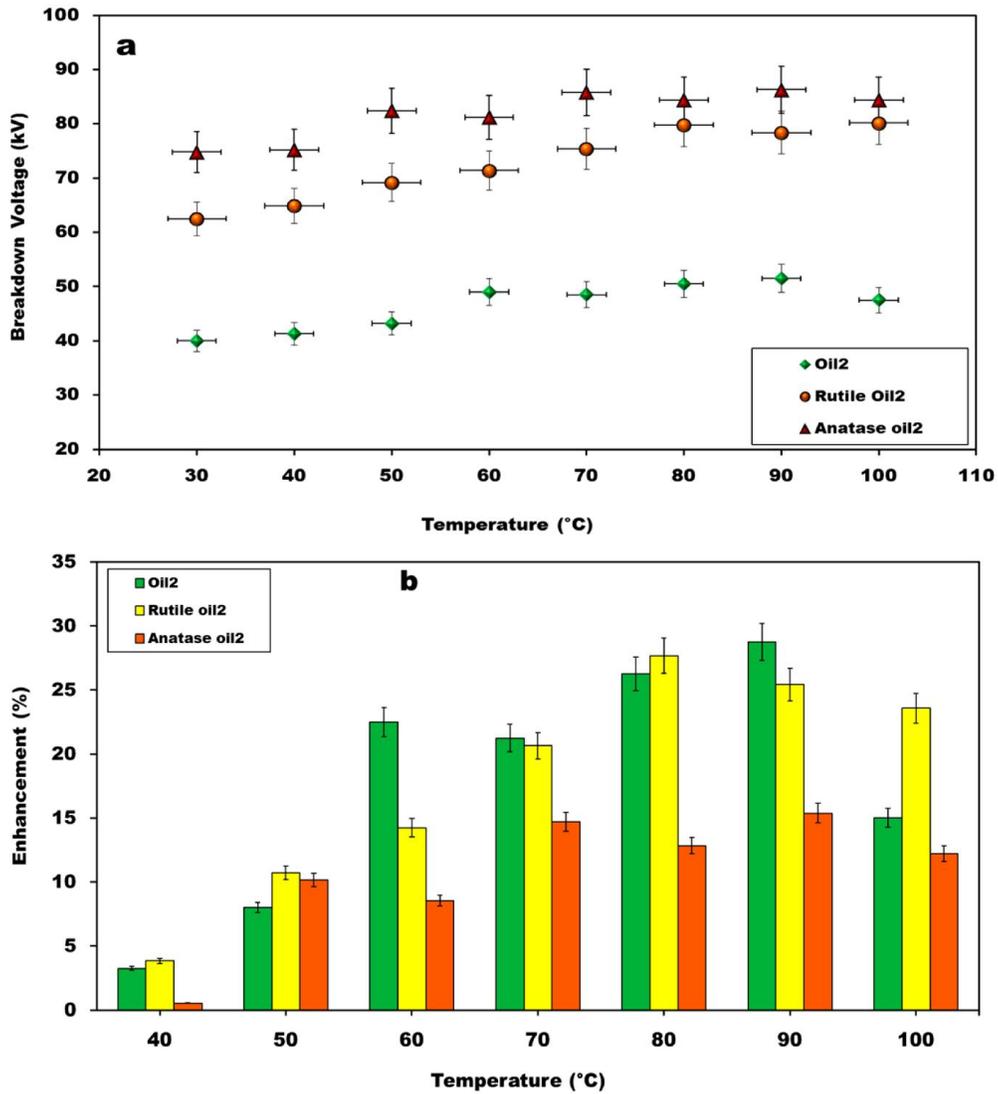

**Figure 11: (a)** Effect of temperature on the dielectric breakdown behaviour of the N.O.s (Anatase oil 2 and Rutile oil 2) with respect to oil temperature. **(b)** The corresponding magnitudes of percentage enhancement for BD voltage for the same, as compared to the performance at room temperature.

### 3.3 Influence of moisture content on the dielectric performance of insulating oils

Presence of moisture in the T.O has been reported to significantly affect the dielectric BD strength of oil i.e., the dielectric strength reduces and failure of the electrical device occurs at lower field strengths. Thereby, the effect of moisture on the performance of N.O.s is a requirement for utility in real systems. In order to comprehend the effect of moisture on BD performance of N.O.s, the moisture level is varied from ~ 20–55 ppm and deterioration in BD strength for the same at critical particle concentration is measured and has been illustrated in Fig. 12 (a). The corresponding magnitude of deterioration in BD strength has been illustrated in Fig. 12 (b). The moisture content has been varied by leaving the N.O.s undisturbed in the ambience over a period of time, during which gains moisture from the atmosphere. It is observed that increment in the amount of moisture present in the oil reduces the BD strength of N.O.s. Experimentally it has been found that moisture level range of 20–25 ppm has insignificant effects on the DB strength but when moisture level varies from 35–55 ppm, the same reduces drastically.

The physics behind the deterioration of BD performance due to the presence of moisture can be explained based on the phenomena of microscopic bubble formation, wherein the surfaces of the bubbles commencing from current impulses onto an electrode gets impregnated with charge carriers during the next impulse. These charge carrier micro bubbles lead to current augmentations and finally breakdown occur earlier [42]. Enhanced moisture content acts a source of increased bubble formation due to vaporization of the water molecules during impulse arc discharges, leading to additional carrier population. The maximum deterioration is found to be ~ 55.6, 31, and 34 %, for T.O, Rutile oil 1 and Anatase oil 1 respectively. Figure 13 (a) illustrates the magnitudes of deterioration in BD voltage as a consequence of moisture content in the oils. It is observed that oil 2, Rutile oil 2 and Anatase oil 2 demonstrate the maximum deterioration in BD strength as ~ 28.2, 38 and 32 %, respectively (Fig. 13 (b)). The reason for reduced deterioration due to presence of nanostructures can be attributed to the moisture affinity of the nanostructures, as observed during storage. The nanostructures imbibe certain quantity of the total moisture content onto themselves by surface sorption, thereby reducing the amount of water molecules available for undergoing vaporization and consequent charge transport.

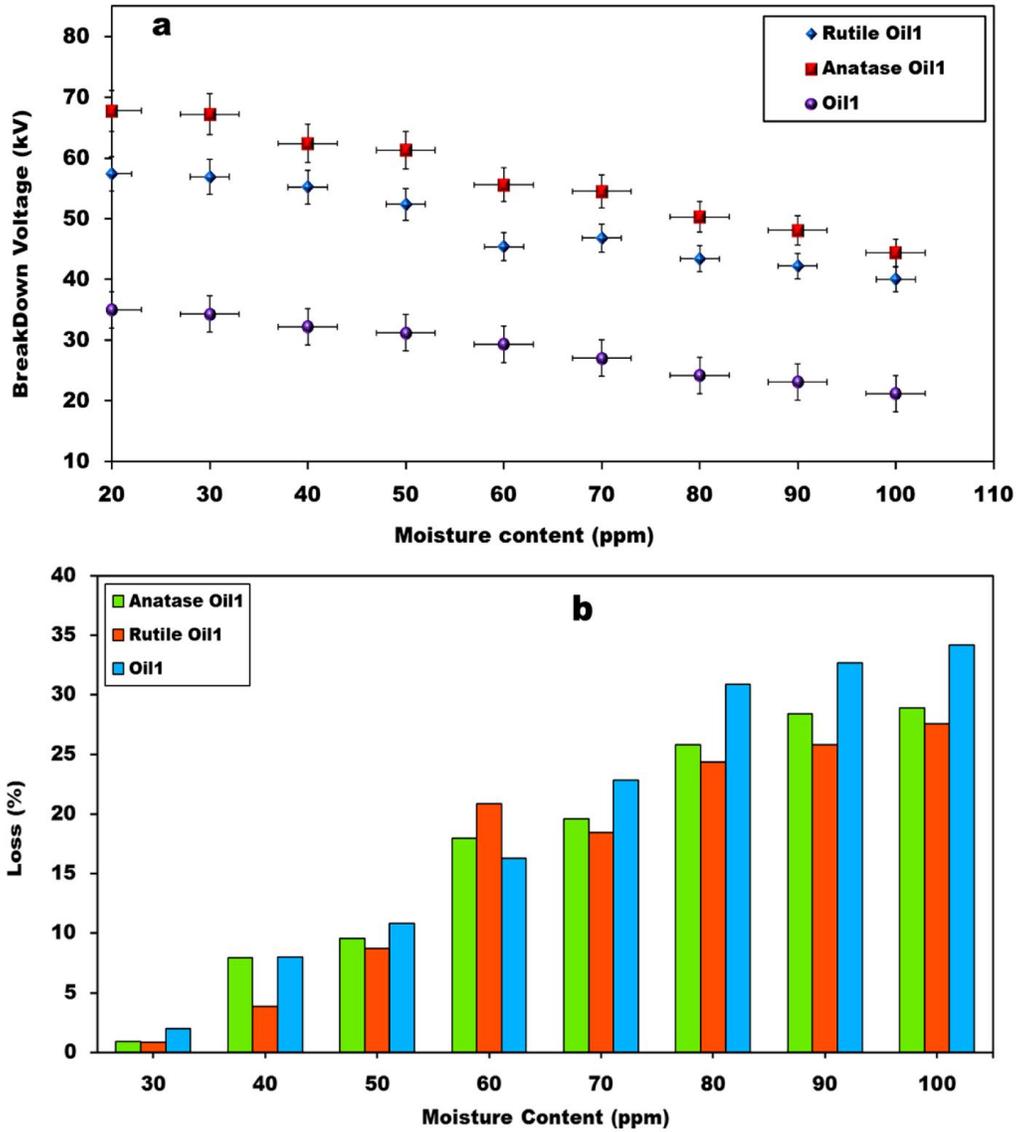

**Figure 12: (a)** The variation of BD voltage as a function of moisture content for N.O.s (oil 1, Anatase oil 1 and Rutile oil 1). **(b)** The respective magnitudes of percentage loss in BD voltage for the same as a consequence of increasing moisture content.

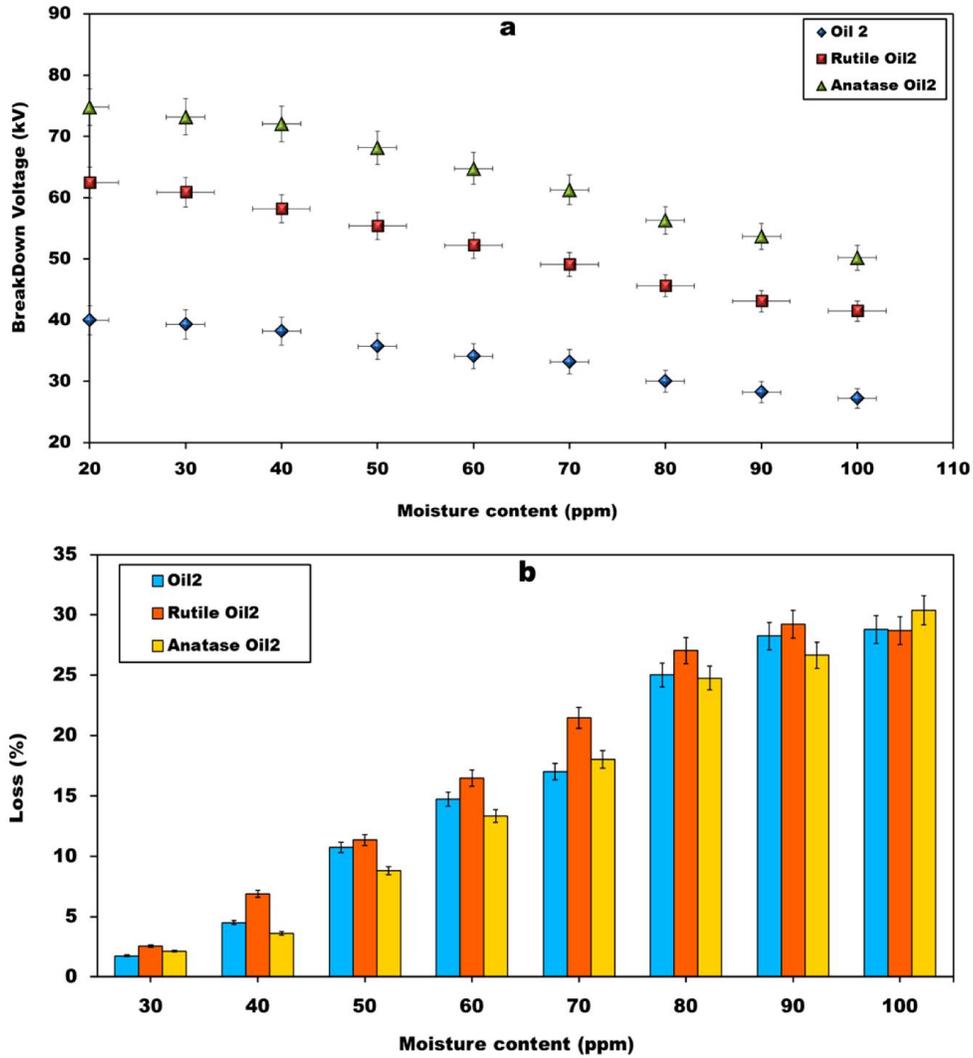

**Figure 13:** **(a)** BD voltage, as a function of moisture content for different oils (i.e. oil 2, Anatase oil 2 and Rutile oil 2). **(b)** The respective magnitudes of percentage loss in BD voltage for the same as a consequence of increasing moisture content.

# 4. Conclusions

The present article reports the utility of $TiO_2$ nanostructures as agents that enhance the dielectric breakdown performance of T.Os. $TiO_2$ (anatase and rutile) infused N.O.s have been formulated by dispersing minute amounts of the nanostructures in two grades of T.O., oil 1 and oil 2, and the dielectric performance has been experimentally studied. Large scale augmentation in the BD strength has been observed at dilute concentrations of NPs for all the tested samples compared to the base oils. The anatase N.O. demonstrates high BD strength compared to the rutile N.O.s but at relatively higher concentration, and the same has been explained based on charge scavenging dynamics and the differences of electronic structure of the two forms of nanomaterial. The BD strength augmentation of the N.O.s at elevated temperatures of those mimicking operational conditions has also been presented. However, it is found that the presence of moisture in the oils reduces the BD performance of NOs (if temperature is maintained constant) accordingly, but by smaller degrees than that observed for the base oils. The present work finds the possibility of application of $TiO_2$ nanostructures in enhancing the operational life of high voltage power systems.

## Acknowledgements

The authors thank the staff of Sophisticated Analytical Instruments Facility (SAIF), IIT Madras for material characterizations. AK is also thankful to Dr. V. Ramanujachari, Director, Research and Innovation Centre (DRDO), Chennai, for permission to publish the present work and for technical inputs. AK also thanks Mr. Manoj Kumar Bhardwaj, Senior Technical Assistant, Fuel and Lubricant Division, DMSRDE Kanpur, for helping in characterization of base oils. The authors acknowledge the financial support by the Defence Research and Development Organization (DRDO) of India (Grant no. ERIP/ER/RIC/2013/M/ 01/2194/D (R&D)). PD would also like to thank IIT Madras for the Pre-doctoral Fellowship.